# Epitaxial growth and electronic structure of Ruddlesden-Popper nickelates (La$_{n+1}$Ni$_n$O$_{3n+1}$, $n$=1-5)

Z. Li[1,2], W. Guo,[1,2] T. T. Zhang,[1,2] J. H. Song,[1,2] T.Y. Gao,[1,2] Z. B. Gu[1,2] and Y. F. Nie[1,2,a]

[1]National Laboratory of Solid State Microstructures, Jiangsu Key Laboratory of Artificial Functional Materials, College of Engineering and Applied Sciences, Nanjing University, 210093 Nanjing, China

[2]Collaborative Innovation Center of Advanced Microstructures, Nanjing University, 210093 Nanjing, China

a) Author to whom correspondence should be addressed: ynie@nju.edu.cn

## ABSTRACT

We report the epitaxial growth of Ruddlesden-Popper nickelates, La$_{n+1}$Ni$_n$O$_{3n+1}$, with $n$ up to 5 by reactive molecular beam epitaxy (MBE). X-ray diffractions indicate high crystalline quality of these films and transport measurements show strong dependence on the $n$ values. Angle-resolved photoemission spectroscopy (ARPES) reveals the electronic structure of La$_5$Ni$_4$O$_{13}$, showing a large hole-like pocket centered around the Brillouin zone corner with a ($\pi$, $\pi$) back-folded copy.

## I. INTRODUCTION

Layered perovskite nickelates have long been considered as close analogs to high T$_c$ cuprates and have been extensively studied by theoretical and experimental investigations for potential new superconductors [1,2,3]. Recently, remarkable progress has been achieved by Li and his collaborators in the synthesis and observation of superconductivity in Sr doped infinite-layer nickelate NdNiO$_2$[4,5], which has been reproduced by other groups [6,7]. NdNiO$_2$ is synthesized by removing the apical oxygen atoms of the perovskite NdNiO$_3$ through a careful reduction process employing CaH$_2$[4,6].

In addition to the infinite-layer compound, the $n=\infty$ member of the Ruddlesden-Popper (RP) phases ($Ln_{n+1}$Ni$_n$O$_{3n+1}$, $Ln$: rare earth elements), the other RP members are natural candidates to be explored for high T$_c$ superconductivity. The homologous series of high T$_c$ cuprates exhibits a clear dependence of the superconducting transition temperature with number of layers in their layered structure [8,9,10], making it more tempting to investigate the RP nickelates with higher $n$ values. In RP structure (Fig. 1), each $n$ layers of corner-shared perovskite octahedra are separated by a $RO$-$RO$ rock-salt structure, which breaks the bonding between adjacent perovskite slabs. By varying the $n$ value, the crystalline structure of $R_{n+1}Ni_nO_{3n+1}$ can be tuned from two-dimensional ($n$=1) to three-dimensional ($n$=∞) with nominal valence state of Ni from $Ni^{2+}$ to $Ni^{3+}$ and $d$ orbital occupation from $d^7$($n$=∞) to $d^8$ ($n$=1). However, these are far from the nearly $d^9$ state in cuprates. By using reduction methods to complete remove the oxygen atoms from the LaO layers in the perovskite blocks and rearrange the oxygen atoms in $(LaO)_2$ blocks[11,12], the RP phases can be turned into the square-planar phases ($R_{n+1}Ni_nO_{2n+2}$) with average $d$ orbital occupation ranging from $d^8$($n$=1) to $d^9$($n$=∞)[13]. Starting from $NdNiO_2$, the reduction product of the $n$=∞ RP member, superconducting dome spanning $d^{8.75}$- $d^{8.875}$ was found by substituting $Nd^{3+}$ with $Sr^{2+}$[4, 5, 6]. Since the valence state and $d$ orbital occupation can also be tuned by only changing the $n$ value, giving the opportunity to realize the optimal doping even without cation substitutions (Fig. 1). The RP phases with high $n$ values are believed to be more promising as cuprate analogs since it has large orbital polarization[14] and the $T_c$ may be enhanced at certain $n$ values like that in cuprates[8,9,10]. Moreover, these reduced RP phase nickelates have many other interesting properties such as dimensionality-controlled insulator-metal transition [15], stripe order [16], magnetic order [17], spin stripe order [18], charge order [19,20], magnetic order-disorder transitions [21], *etc*.

Up to date, however, the RP phase with highest $n$ value (except perovskite, the $n$=∞ member) is $La_4Ni_3O_{10}$ ($n$=3). The corresponding square-planar phase ($La_4Ni_3O_8$) has a $Ni^{1.33+}$ ($d^{8.67}$) configuration [12,22], which is still out of the superconducting dome of cuprates and nickelates. As such, synthesizing RP nickelates with high $n$ values to explore the potential superconductivity in layered nickelates is highly desired, but it is rather challenging since these structures are metastable and only achievable by using a layer-by-layer deposition techinque [23].

Here we report the synthesis of RP nickelate series ($La_{n+1}Ni_nO_{3n+1}$) with $n$ values up to 5 using MBE. X-ray diffraction indicates the high crystalline quality of these films and transport measurements show a strong dependence on the $n$ value. The electronic structure of the $n$=4 member is also investigated.

## II. MBE GROWTH, STRUCTURE AND TRANSPORT MEASUREMENTS OF $La_{n+1}Ni_nO_{3n+1}$ THIN FILMS

RP $La_{n+1}Ni_nO_{3n+1}$ thin films were epitaxially grown on (001) $LaAlO_3$ (LAO) single-crystalline substrates from MTI at 550 ℃ and ~1.5× $10^{-5}$ Torr of distilled ozone using a DCA R450 MBE system. Before growth, the LAO substrates were dropped in boiling water for 15 minutes, then annealed in air furnace for 600 minutes at 1000℃ and dropped in boiling water again to obtain an $AlO_2$-terminated flat surface, similar to the methods previously employed in the literature [24]. During growth, reflection high-energy electron diffraction (RHEED) with a 15 keV electron beam was employed to monitor and control the growth process and film quality. The film crystalline structure was examined by high-resolution x-ray diffraction (XRD) using a Bruker D8 Discover diffractometer. The transport properties of films were measured using Van der Pauw method by a Quantum Design PPMS system and the electrical contacts were made using ultrasonic wire bonding.

To obtain high quality layered RP epitaxial films, especially the non-thermal equilibrium phase with high $n$ values ($n >3$), a layer-by-layer deposition using shutter-control method is needed. In addition to the requirement of correct stoichiometry (La:Ni ratio), precise monolayer dosage of LaO and NiO$_2$ layers are also extremely critical for the synthesis of high crystalline quality RP films, especially for high $n$-value RP members that the imperfect layer dosage cannot be accommodated through a self-resembled process. Typically, a shutter-control layer-by-layer deposition of SrTiO$_3$, BaTiO$_3$ and many other perovskite oxides exhibit a nearly perfect sinusoidal curve pattern. The diffraction intensity saturates near the end of the growth of a full atomic monolayer, providing clear signatures for optimizing the shutter time for each source. However, the RHEED patterns of LaNiO$_3$, LaAlO$_3$ and many other La-related perovskites do not show ideal sinusoidal curve patterns in a shutter-control growth mode, which precludes the calibration of monolayer dosages with high precision. Instead, precise shutter times for La and Ni sources were calibrated by optimizing the growth parameters of the simplest RP member, LaNiO$_3$, using a co-deposition method [25]. After a rough calibration of La and Ni flux ratio to be 1:1 using a quartz crystal microbalance (QCM), the shutter of La and Ni sources were opened to deposit LaO and NiO$_2$ layers simultaneously.

During the calibration process, the intensity of (11) diffractions in the RHEED patterns shows a clear oscillation pattern (Fig. 2a) and each period of the oscillations is corresponding to the growth of about one unit cell. In addition to the oscillations, the overall average intensity is sensitive to the film stoichiometry and drifts to higher (lower) intensity when Ni (La) is rich. By adjusting the source temperature to optimize the beam flux ratio to be 1:1, a stable oscillation pattern can be obtained and yield precise monolayer dosages for LaO and NiO$_2$ layers. By extracting the oscillation period from more than 10 oscillation cycles, precise shutter times for La and Ni sources were obtained. Based on these monolayer dosages calibrated by a co-deposition method, RP nickelates of any $n$ value can be synthesized by deposing LaO and NiO$_2$ layers alternatively in the desired sequences using a shutter-control mode. During the shutter-control growth process, the RHEED electron beam was closed and the substrates were rotating to increase the uniformity of the films.

As shown in Fig. 2b, high resolution XRD measurements confirm the crystalline quality of the nickelate films. All films exhibit clear and sharp (00L) diffraction peaks in the $2\theta$-$\omega$ scans for all superlattice peaks even the films are rather thin (only 20 formula units for LaNiO$_3$ and 5 formula units for all other RP films). This indicates that the combination of co-deposition calibration and shuttered-growth is an effective and precise method to grow RP phase nickelates. All the films are fully strained to the substrate and a representative reciprocal space mapping (RSM) of $n$=4 compound is shown in Fig. 2d. The extracted $c$ lattice constants are listed in Table I. The LaNiO$_3$ films are atomically smooth showing clear step and terrace feature in the atomic force microscopy (AFM) image while the RP phases with (LaO)$_2$ are less smooth and only show weak step and terrace feature (Fig. 2c). Nonetheless, the crystalline quality of the RP films are also of high quality. Although we here only demonstrate the growth of RP phases with *n* up to 5, members with higher *n* values are expected to be able to be synthesized using this method.

Table I. Out-of-plane lattice constants of Ruddlesden-Popper La$_{n+1}$Ni$_n$O$_{3n+1}$ films grown on LaAlO$_3$ substrates

| *n* value | 1 | 2 | 3 | 4 | 5 | ∞ |
|---|---|---|---|---|---|---|
| *c* lattice constant (Å) | 12.807 | 20.705 | 28.284 | 36.085 | 43.767 | 3.887 |

Transport properties of La$_{n+1}$Ni$_n$O$_{3n+1}$ films show a clear dependence on the *n* value (Fig. 3). The *n*=1 member is fully insulating at room temperature, which is consistent with the literature. The *n*=2 compound is sensitive to the detailed crystalline structure or the oxygen stoichiometry and whether it is metallic or insulating is still controversial.[26-34] In our epitaxial thin film case, the higher crystalline quality *n*=2 compound exhibits insulating temperature dependence. We also notice that the crystalline quality has strong impact on the transport property. Some *n*=2 films exhibit broad x-ray diffraction peaks and metallic behavior (not shown). For the *n*=3 compound, our films show a metallic behavior down to the lowest measured temperature and do not show any upturn or kink (signature of ordering) as reported previously.[26,28,30,32,33,34] It is most likely that the compressive epitaxial strain imposed by the substrate increases the electron hopping integral and suppresses the gap opening. For *n*≥3, La$_{n+1}$Ni$_n$O$_{3n+1}$ are also metallic and the resistivity monotonically decreases as n increases, which could be explained by the higher electron hopping integral due to the increase of the dimensionality and the compressive strain imposed by the substrate.

### III.  ELECTRONIC STRUCTURE OF La$_5$Ni$_4$O$_{13}$ films

Following the growth, La$_5$Ni$_4$O$_{13}$ films were transferred through an ultrahigh vacuum (<1×10$^{-10}$ Torr) into a high-resolution ARPES system with a VG Scienta R8000 electron analyzer and a VUV5000 helium plasma discharge lamp. ARPES measurements were performed using He Iα (hν = 21.2 eV) photons with an energy resolution ΔE of 11 meV. As shown in Fig. 4, ARPES measurements were performed on La$_5$Ni$_4$O$_{13}$ at 9K. The Fermi surface shows a gapless large hole-like pocket centered at the Brillouin zone corner, which is similar to the LaNiO$_3$[25] and La$_4$Ni$_3$O$_{10}$[35] and similar to the other nickelates[25,错误!未定义书签。] and closely resembles the Fermi surface of optimally hole-doped cuprates[36]. The electron-like band centered around the zone center observed in La$_4$Ni$_3$O$_{10}$ [35] is not visible in this case. This is most likely due to the strong k$_z$ dispersion of this band that cannot be probed by He Iα photons, similar to the LaNiO$_3$ case [25]. The original data of the E vs K cuts are relatively vague and the band dispersion can be clearly visualized by subtracting the background as shown in Fig.4c and Fig.4d. The hole-like band shows a dispersive feature with the band bottom located at around 130 ± 4meV as extracted from a parabolic fit to the E vs K cut along the (0.5π, 0) direction. The cut along the (π, 0) direction also shows a hole-like band crossing the Fermi level Clearly the hole-like band crosses through the Fermi level and show no gap around the whole momentum space, similar to that in LaNiO$_3$[25] and La$_4$Ni$_3$O$_{10}$[35] but distinct from the pseudogap feature observed in $R_{2-x}$Sr$_x$NiO$_4$ (R=Nd, Eu)[37]. Moreover, there exist a clear ($\pi$, $\pi$) back-folded copy of the hole-like band (marked by dotted green line), which is similar to cuprates[38,39,40,41] and La$_4$Ni$_3$O$_{10}$[35] but different from LaNiO$_3$[25]. This band folding is consistent with the superlattice diffractions observed in the low energy electron diffraction (LEED) pattern (Fig. 4a) and most likely due to the lattice reconstruction driven by NiO$_6$ octahedral rotations. Also, the energy gap and kink feature observed in the $n$=3 compound[35] are absent in our ARPES and transport measurements on the $n$=4 compound, which may be due to the enhancement of the electron hopping integral due to the increase of dimensionality and compressive strain imposed by the substrate.

## VI. DISCUSSION AND CONCLUSION

One of the main interests in RP nickelates is to engineer the $d$ orbital occupation for high T$_c$ superconductivity. By removing the oxygen atoms from the LaO layers in the perovskite blocks and rearrange the oxygen atoms in (LaO)$_2$ blocks in La$_5$Ni$_4$O$_{13}$ and La$_6$Ni$_5$O$_{16}$, the average Ni valance states in the reduced phases of La$_5$Ni$_4$O$_{10}$ and La$_6$Ni$_5$O$_{12}$ are Ni$^{1.25+}$ and Ni$^{1.2+}$ with an average 3$d$ filling of $d^{8.75}$ and $d^{8.8}$, respectively. With no need for cation substitution, these layered nickelates are within the superconducting dome in cuprates[36] and infinite-layer nickelates[4,6]. Such reduction process has been applied in the synthesis of LaNiO$_2$, NdNiO$_2$ and La$_3$Ni$_2$O$_7$ by annealing in Ar/H$_2$ atmosphere[16] or vacuum sealed with CaH$_2$ powder[42]. However, the reduction process is sensitive to many factors and very challenging to obtain superconducting phases. The reduction of these films needs to be further explored in the future and is out of the scope of this work. Also, as Sr doped LaNiO$_2$ is not superconducting [4], whether the other La-based RP series would be superconducting is an interesting question to be investigated. Moreover, the synthesis method for La-based RP series demonstrated in this work can be applied in Nd-based RP series.

In summary, we synthesize a series of RP lanthanum nickelate thin films with $n$ up to 5 by using reactive MBE. Especially, the metastable $n$=4 and $n$=5 compounds are synthesized for the first and only achievable by using a layer-by-layer deposition method. A combination of co-deposition and shuttered growth methods were employed to achieve precise monolayer-precise control, allowing the synthesis of high quality RP members with high $n$ values. Transport measurements show an increase of the conductivity with the $n$ value. The La$_5$Ni$_4$O$_{13}$ films show a large hole-like pockets centered at zone corner with a ($\pi$, $\pi$) back-folded copy, similar to high T$_c$ cuprates and other layered nickelates. The synthesis of RP nickelates with high $n$ values provide a great platform to explore high T$_c$ superconductivity upon further reduction and charge carrier doping.

## ACKNOWLEDGMENTS

This work was supported by the National Natural Science Foundation of China (Grant Nos. 11774153, 11861161004, 51672125, and 51772143) and the Fundamental Research Funds for the Central Universities (Grant No. 0213-14380167).

**Availability of data**：The data that support the findings of this study are available from the corresponding author upon reasonable request.

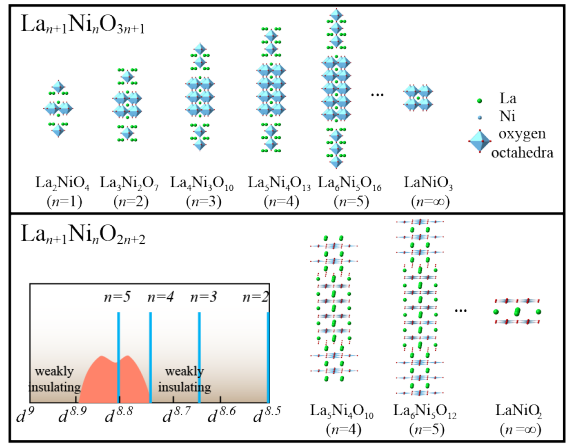

FIG.1 Crystal structures of La$_{n+1}$Ni$_n$O$_{3n+1}$ and La$_{n+1}$Ni$_n$O$_{2n+2}$. Insert shows the phase diagram of Sr doped infinite layer nickelates reported in Ref.[5,6] and the 3$d$ orbital occupation of square-planar phase La$_{n+1}$Ni$_n$O$_{2n+2}$ with different $n$ values.

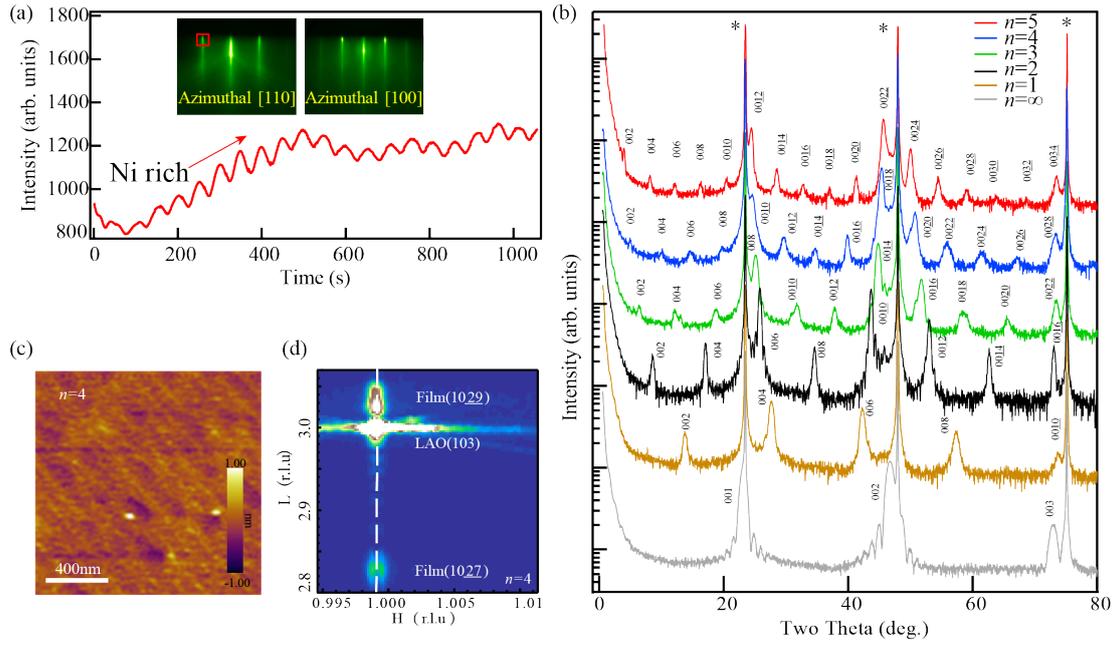

FIG. 2. (a) RHEED patterns and RHEED intensity oscillations by monitoring the 11 diffraction (red square) during the growth of LaNiO$_3$. The average RHEED intensity increases when Ni is rich and stabilizes with a flux ratio of La:Ni=1:1. (b) XRD $2\theta$-$\omega$ scan patterns of Ruddlesden-Popper LNO films grown on (001) LaAlO$_3$ substrates. The thickness of LaNiO$_3$ is 20 formula units and the thickness of other RP films are all 5 formula units. The asterisks denote diffraction peaks from the substrate. (c) AFM image of the surface morphology of a 5 u.c. La$_5$Ni$_4$O$_{13}$/LaAlO$_3$ (001) sample. (d) Reciprocal space map around the 103 diffraction peak of the substrate.

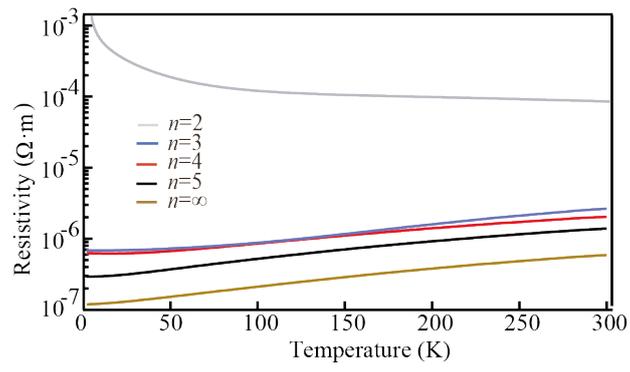

FIG.3. Resistivity v.s. temperature curves of $La_{n+1}Ni_nO_{3n+1}$ films. The $n$=1 films are insulating and out of the measurement limit.

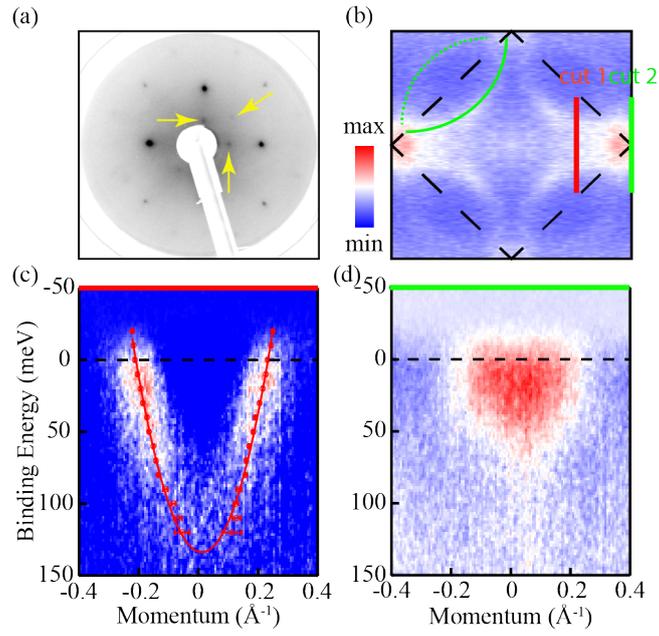

FIG4. (a) LEED pattern taken with an incident electron beam energy of 100 eV for La$_5$Ni$_4$O$_{13}$ films showing superlattice diffractions (pointed by yellow arrows). (b) Symmetrized Fermi surface of first Brillouin zone showing (π, π) back-folded large hole-like pocket centered at the zone corner. The dashed lines represent the folded Brillouin zone. (c, d) E vs K maps taken along cut I and cut II shown in panel (b).